\documentclass{aastex61}
\usepackage{amsmath}
\usepackage{tabularx}
\usepackage{tabularht}
\usepackage{rotating}
\usepackage[colorinlistoftodos]{todonotes}
\usepackage{hyperref}
\hypersetup{linkcolor=red,citecolor=blue,filecolor=cyan,urlcolor=magenta}
\usepackage[hyphenbreaks]{breakurl}
\usepackage[title]{appendix}

\usepackage{multirow}

\shorttitle{Statistical Relationship Between LDGRFs and SEPs}
\shortauthors{Bruno et al.}

\begin{document}
\title{Statistical Relationship Between Long-duration High-Energy Gamma-Ray Emission and Solar Energetic Particles}

\author{A.~Bruno}\affiliation{Heliophysics Division, NASA Goddard Space Flight Center, Greenbelt, MD, USA.}
\affiliation{Department of Physics, Catholic University of America, Washington DC, USA}
\author{G.~A.~de~Nolfo}\affiliation{Heliophysics Division, NASA Goddard Space Flight Center, Greenbelt, MD, USA.}
\author{J.~M.~Ryan}\affiliation{Space Science Center, University of New Hampshire, Durham, NH, USA.}
\author{I.~G.~Richardson}\affiliation{Heliophysics Division, NASA Goddard Space Flight Center, Greenbelt, MD, USA.}\affiliation{Department of Astronomy, University of Maryland, College Park, MD, USA.}
\author{S.~Dalla}\affiliation{Jeremiah Horrocks Institute, University of Central Lancashire, Preston PR1 2HE, UK.}

\correspondingauthor{A. Bruno}
\email{alessandro.bruno-1@nasa.gov}

\begin{abstract}
Large solar eruptions are often associated with long-duration $\gamma$-ray emission extending well above 100 MeV. While this phenomenon is known to be caused by high-energy ions interacting with the solar atmosphere, the underlying dominant acceleration process remains under debate. Potential mechanisms include continuous acceleration of particles trapped within large coronal loops or acceleration at coronal mass ejection (CME)-driven shocks, with subsequent back-propagation towards the Sun. As a test of the latter scenario, previous studies have explored the relationship between the inferred particle population producing the high-energy $\gamma$-rays, and the population of solar energetic particles (SEPs) measured in situ. However, given the significant limitations on available observations, these estimates unavoidably rely on a number of assumptions. In an effort to better constrain theories of the $\gamma$-ray emission origin, we re-examine the calculation uncertainties and how they influence the comparison of these two proton populations. We show that, even accounting for conservative assumptions related to $\gamma$-ray flare, SEP event and interplanetary scattering modeling, their statistical relationship is only poorly/moderately significant. However, though the level of correlation is of interest, it does not provide conclusive evidence for or against a causal connection. The main result of this investigation is that the fraction of the shock-accelerated protons required to account for the $\gamma$-ray observations is $>$20--40\% for six of the fourteen eruptions analyzed. Such high values argue against current CME-shock origin models, predicting a $<$2\% back-precipitation, hence the computed numbers of high-energy SEPs appear to be greatly insufficient to sustain the measured $\gamma$-ray emission.
\end{abstract}

%
%
%
%

\section{Introduction}\label{Introduction}
High-energy ($>$100 MeV) photons in
long-duration $\gamma$-ray flares (LDGRFs) are known to originate from the decay of pions produced in the interaction of $>$300 MeV protons and $>$200 MeV/n $\alpha$ particles with the solar chromosphere and photosphere (e.g., \citealp{ref:VILMER2011}).
LDGRFs are characterized by a delayed and prolonged emission, extending up to tens of hours after the impulsive phase \citep{ref:RYAN2000}. Despite the significant progress in recent years as a result of the observations of the $Fermi$ Large Area Telescope (LAT) (see \citealp{ref:AJELLO2021} and references therein), the
involved processes are still controversial.
Two main competing scenarios have been proposed to explain
the acceleration of the interacting particles responsible for 
the origin of LDGRFs and their subsequent precipitation into the solar atmosphere:
1) particle trapping with/without continuous acceleration within large ($\gtrsim$1 R$_{\odot}$) coronal loops, characterized by a delayed onset representing the time required by the ions to exceed the pion production threshold energy \citep{ref:RYANLEE1991,ref:MANDZHAVIDZERAMATY1992,ref:CHUPPRYAN2009,ref:GRECHNEV2018,ref:RYAN2018,ref:DENOLFO2019,ref:RYAN2019,ref:DENOLFO2021}, and 2) coronal mass ejection (CME)-driven shock acceleration, i.e. the dominant mechanism for gradual solar energetic particle (SEP) events measured in situ
\citep{ref:WILD1963,ref:RAMATY1987,ref:CLIVER1993,ref:KOCHAROV2015,ref:PESCE2015,ref:PLOTNIKOV2017,ref:JIN2018,ref:KAHLER2018,ref:GOPALSWAMY2018,ref:KOULOUMVAKOS2020}. In the latter case, the high-energy photon emission is also referred to as late-phase $\gamma$-ray emission \citep{ref:SHARE2018} or sustained $\gamma$-ray emission \citep{ref:KAHLER2018}.
Alternative models predict particle trapping/(re-)acceleration in non-flaring closed loops \citep{ref:HUDSON2018} or by electric fields in a current sheet in the wake of a CME (e.g., \citealp{ref:AKIMOV1996,ref:KOCHAROV2020}).

While LDGRFs tend to be associated with relatively fast CMEs and large SEP events, often with energies typical of ground-level enhancements (GLEs), the apparent connection between these phenomena is questioned by noteworthy
counter-examples.
For instance, LDGRFs accompanied by CMEs with space speeds as low as 830 km s$^{-1}$ have been reported, based on the Coordinated Data Analysis Workshop (CDAW) catalog;
in addition, the mean velocity for the sample of 35 LDGRFs listed by 
\citet{ref:DENOLFO2019} was $\sim$1705 km s$^{-1}$ ($\sim$1732 km s$^{-1}$ excluding the two partial-halo CMEs), with 40\% of them slower than 1500 km s$^{-1}$. For comparison, the average speed of CMEs associated with the 23$^{rd}$ solar cycle GLEs was 2083 km s$^{-1}$ \citep{ref:GOPALSWAMY2012}.
Of particular note is that no CME was observed during the events on 2012 October 23 and November 27, 
suggesting that a fast and wide CME is not a necessary condition for LDGRFs,
although there is evidence for the eruption of magnetic loops and material moving away from the flare region \citep{ref:SHARE2018}. On the other hand, 
the 2012 May 17 GLE event was not linked to one of the larger LDGRFs, and
some of the high-energy SEP events measured by the \textit{Payload for Antimatter Matter Exploration and Light-nuclei Astrophysics} (PAMELA) space experiment were not associated with LDGRFs;
analogously,
several LDGRFs were accompanied by relatively small SEP events \citep{ref:DENOLFO2019}. Finally, while characterized by one of the longest $\gamma$-ray emissions, no significant flux of high-energy SEPs was measured during the 2011 March 7 event;
furthermore, the particle release time inferred from L1 observations occurred at least $\sim$10 minutes later than the LDGRF onset \citep{ref:KAHLER2018,ref:KLEIN2018}. 

The possible link between the populations of shock-accelerated ions escaping into  interplanetary space and those precipitating onto the solar atmosphere was explicitly investigated by \citet{ref:SHARE2018} and, more thoroughly, by \citet{ref:DENOLFO2019}, who used the high-energy proton observations from PAMELA, the \textit{Geostationary Operational Environmental Satellites} (GOES) and the twin \textit{Solar TErrestrial RElations Observatory} (STEREO) spacecraft to reconstruct the spatial distribution of fourteen SEP events associated with LDGRFs. The resulting numbers of $>$500 MeV protons at 1 AU ($N_{SEP}$) were compared to the corresponding numbers of $>$500 MeV protons producing the LDGRFs ($N_{LDGRF}$), as inferred from the $Fermi$-LAT $\gamma$-ray observations by \citet{ref:SHARE2018}. No correlation was found between the $N_{SEP}$ and the $N_{LDGRF}$ values, suggesting that the two phenomena are not produced by the same population of CME-driven shock accelerated ions. 
Furthermore, the corresponding precipitation fractions $P_{N}$ = $N_{LDGRF}$ / ($N_{LDGRF}$ + $N_{SEP}$) were found to be characterized by large values, with eight (six) events having $P_{N}$ $>$ 10\% (20\%) and even three events with $P_{N}$ $>$ 80\%, implying that a very large percentage of the overall shock-accelerated particle population would be required to explain the observed LDGRF fluence.
These large precipitation fractions are clearly inconsistent with current model predictions, which suggest that SEP back-propagation from 
the height of the CME-driven shock down
to the solar surface is strongly impeded by magnetic mirroring \citep{ref:HUDSON2018,ref:KLEIN2018}. In fact, only ions injected nearly parallel to the coronal or interplanetary magnetic field lines 
near the shock, in a narrow loss cone, can reach a sufficiently dense region of the solar atmosphere to undergo nuclear interactions. Assuming an isotropic particle distribution at the shock, the corresponding fraction amounts to $\sim$1\% of the initial population \citep{ref:KLEIN2018}.
Recently, \citet{ref:HUTCHINSON2022} investigated the mirroring problem extensively using 3D test particle simulations with varying levels of scattering. While strong scattering conditions can occur close to the Sun, back-precipitation
was shown to be generally highly inefficient, with instantaneous precipitation fractions lower than 2\%. In addition, a strong radial dependence was found for $P_{N}$, so that the back-precipitation drastically decreases with increasing injection heights as the CME shock expands.
An upper limit on the total precipitation fraction in the CME scenario was evaluated for eight of the fourteen events analyzed by \citet{ref:DENOLFO2019}, with values ranging from $\sim$0.56\% to $\sim$0.93\%, increasing with decreasing shock speed \citep{ref:HUTCHINSON2022}. In fact, faster shocks accelerate particles over larger helio-distances, and tend to spend less time close to the solar surface, where the precipitation is more efficient. 
These modeled precipitation fractions
are, on average, almost a factor $\sim$50 smaller than the values 
reported by \citet{ref:DENOLFO2019}.
In addition, the precipitation temporal profiles assessed with the simulations exhibit a much faster decay with respect to experimental observations. Thus, 
although solar protons are believed to be accelerated to the requisite energy range by CME-driven shocks, their transport in sufficient numbers back to the Sun by means of a robust, repeatable  process is a challenge for the widely invoked CME-shock origin model for LDGRFs.

In this work we critically analyze the assumptions behind the calculation of the $N_{LDGRF}$ and $N_{SEP}$ numbers, and further explore the statistical relationship between 
LDGRFs and SEP events to test the validity of the CME-shock paradigm.
The article is structured as follows. In Section \ref{Interacting and interplanetary Protons} we analyze the uncertainties affecting the comparison between the precipitating and the interplanetary proton populations reported by \citet{ref:DENOLFO2019}, including the effects of $\gamma$-ray flux, SEP event and interplanetary transport modeling, and discuss the comparison with previous calculations. In Section \ref{Gamma-ray and type-II radio emission} we investigate the association between LDGRFs and interplanetary type-II radio emission
that is claimed by \citet{ref:GOPALSWAMY2018b} to support the CME-shock scenario. Finally, Section \ref{Conclusions} summarizes
the study and presents our conclusions.

%
%
%
%
\section{Interacting and interplanetary Protons}\label{Interacting and interplanetary Protons}
%
%
%
%

\subsection{Interacting Protons}\label{Interacting Protons}
Estimates of the numbers of $>$500 MeV protons producing the observed $>$100 MeV $\gamma$-ray emission used in \citet{ref:DENOLFO2019} rely on the work by \citet{ref:SHARE2018}, who carried out a comprehensive analysis of 30 LDGRFs between 2011 and 2015 based on the $Fermi$-LAT data. These authors used the model by \citet{ref:MURPHY1987} to assess the source spatial distribution, accounting for the attenuation effects associated with atmospheric absorption. For instance, their calculation results in $\sim$80\% of photons escaping at a heliocentric angle (or central meridian distance, CMD) of 70$^{\circ}$, with this fraction decreasing to $\sim$47\% and $\sim$8\% at 85$^{\circ}$ and 90$^{\circ}$, respectively. They also estimated the corresponding hardening in the escaping $\gamma$-ray spectrum. In addition, \citet{ref:SHARE2018} provided a heliocentric angle-dependent correction factor, ranging from 1 at solar limb to $\sim$2.3 at disk center,
accounting for the fact that the ions producing LDGRFs may be characterized by an approximately downward isotropic distribution, rather than fully isotropic.
However, while flare-accelerated protons are more likely to follow an approximately downward isotropic distribution, this may not be the case for shock-accelerated protons (see \citealp{ref:SHARE2018} and references therein). Therefore, in contrast to \citet{ref:DENOLFO2019}, the relative correction factor was not used in this analysis, aiming to provide more conservative lower limits on the precipitation fraction.

The $N_{LDGRF}$ numbers from \citet{ref:SHARE2018} were also compared with those recently estimated by 
\citet{ref:AJELLO2021}. The results of the two calculations for the fourteen events analyzed by \citet{ref:DENOLFO2019} are reasonably similar 
with a few exceptions: in particular, according to \citet{ref:AJELLO2021}, $N_{LDGRF}$ is a factor $\sim$1.8, $\sim$2.6 and $\sim$1.5 higher for the 2011 Jun 7, the 2011 September 6 and the 2012 May 17 events, respectively, while it is about a factor $\sim$0.4 for the 2014 September 1 event. 
As discussed below, these differences
are reflected in the corresponding precipitation fraction. 

%
%
%
%

\subsection{Interplanetary Protons}\label{Interplanetary Protons}
The numbers of $>$500 MeV interplanetary protons ($N_{SEP}$) were estimated by \citet{ref:DENOLFO2019} by combining the 
multi-point in-situ observations from PAMELA/GOES, STEREO-A and STEREO-B.
As comprehensively described in \citet{ref:BRUNORICHARDSON2021}, the SEP spatial distribution on the spherical surface with 1 AU radius was derived by means of a 2D Gaussian model, accounting for both longitudinal and latitudinal magnetic connectivity, as a function of the spherical distance:
\begin{equation}\label{eq:great-circle}
\delta = \arccos\left[ \sin(\theta)\sin(\theta_{sep}) + \cos(\theta)\cos(\theta_{sep})\cos(\phi-\phi_{sep})\right],
\end{equation}
where $\theta$ and $\phi$ are the Stonyhurst heliographic latitude and longitude (e.g., of the observing spacecraft footpoint), and ($\theta_{sep}$, $\phi_{sep}$) are the coordinates of the distribution peak (main SEP propagation axis). The distribution standard deviation assumes a spherical symmetry, so that $\sigma^{\theta}$ = $\sigma^{\phi}$ = $\sigma$. The longitude of the SEP propagation axis ($\phi_{sep}$) was obtained by the best fit of the three-spacecraft intensities, while its latitudinal angle $\theta_{sep}$ is not derivable from experimental data.

\begin{table}[!t]
\center
\begin{tabular}{l|c|c|c|c|c|c}
& SEP & Earth & Flare & G2014 & \multicolumn{1}{c|}{DONKI} & \multicolumn{1}{c}{CDAW} \\
No & event & footpoint & location & FR loc. & direction & speed  \\
\hline
1 & 2011/03/07 & S07W55 & N30W48 & N32W58 & N17W50 & 2223  \\
2 & 2011/06/07 & N00W55 & S21W54 & S08W51 & S25W52 & 1321  \\
3 & 2011/08/04 & N05W64 & N19W36 & N19W30 & N14W40 & 1477   \\
4 & 2011/08/09 & N06W42 & N14W69 & N08W68 & S12W62 & 1640   \\
5 & 2011/09/06 & N07W57 & N14W18 & N20W19 & N20W20 & 830   \\
6 & 2012/01/23 & S05W58 & N33W21 & N30W22 & N41W26 & 2511  \\
7 & 2012/01/27 & S05W47 & N33W85 & N27W82 & N40W75 & 2541  \\
8 & 2012/03/07 & S07W62 & N17E27 & N18E31 & N30E60 & 3146  \\
9 & 2012/05/17 & S02W62 & N07W88 & S07W76 & S10W75 & 1596   \\
10 & 2012/07/07 & N03W53 & S13W59 & S29W62 & S35W65 & 1907   \\
11 & 2013/04/11 & S05W66 & N07E13 & N08E11 & S07E25 & 1369   \\
12 & 2013/10/28 & N04W84 & S06E28 & \nodata & N20E10 & 1098   \\
13 & 2014/02/25 & S02W48 & S12E82 & S08E80\tablenotemark{b} & S11E78 & 2153   \\
14 & 2014/09/01 & N02W46 & N14E127\tablenotemark{a} & N16E117\tablenotemark{c} & N01E155 & 2017   \\
\end{tabular}
\tablerefs{
(\textit{a}) \citet{ref:ACKERMANN2017},
(\textit{b}) \citet{ref:GOPALSWAMY2015}, 
(\textit{c}) \citet{ref:GOPALSWAMY2020}
}
\caption{Relevant heliographic coordinates and CME parameters for the fourteen LDGRF-associated SEP events analyzed by \citet{ref:DENOLFO2019}. The first two columns report the SEP event number and date; the third column lists the heliographic coordinates of the Earth footpoint; the fourth column gives the location of the parent flares based on the SolarSoft package; the fifth column indicates the CME flux rope location (not available for event \#12) estimated by \citet[G2014]{ref:GOPALSWAMY2014}; the column 6 reports the CME direction from the DONKI catalog. Finally, column 7
lists the CME space speed (km s$^{-1}$) from the CDAW catalog.}
\label{tab:Table1}
\end{table}

The flare locations and the CME directions/space speeds of the fourteen events analyzed by \citet{ref:DENOLFO2019} are reported in Table \ref{tab:Table1}. 
The first two columns list the event numbers and dates.
The footpoint locations of the magnetic field line passing through the Earth, mapped ballistically back to 2.5 R$_{\odot}$ \citep{ref:BRUNORICHARDSON2021}, are shown in the third column.
Column 4 gives the flare coordinates from the SolarSoft package. The next two columns refer to the CME directions from \citet{ref:GOPALSWAMY2014}, who fitted a flux-rope to the CMEs in the \textit{SOlar and Heliospheric Observatory} (SOHO) and STEREO images using the graduated cylindrical shell (GCS) model \citep{ref:THERNISIEN2011}, and from the Space Weather Database Of Notifications, Knowledge, Information (DONKI) catalog, based
on the geometric triangulation of SOHO and STEREO coronagraph measurements.
Finally, the CME space speeds from the CDAW catalog,
relying on a cone-model correction of sky-plane speeds \citep{ref:GOPALSWAMY2010}, are also reported in the last column.
In general, accounting for the associated uncertainties, the heliographic coordinates of flares and CMEs are relatively similar, while a large deviation suggests a significant non-radial motion component, or that the flare location is not directly under the center of the CME.
On the other hand, there is a remarkable (up to tens of degrees) discrepancy between the CME latitudinal angles in Table \ref{tab:Table1}.
Indeed, it is well-known that current CME catalogs using different analysis methods and different instruments generally disagree on the properties (speed, width and direction) of individual CMEs associated with SEP events \citep{ref:RICHARDSON2015}. However, CME reconstructions based on multiple viewpoints have typically smaller errors (see, e.g., \citealp{ref:VERBEKE2022}). For instance, the uncertainties on the CME latitudinal/longitudinal angles in the DONKI catalog are typically 5$^{\circ}$/10$^{\circ}$ and 15$^{\circ}$/30$^{\circ}$ for parameter estimates based on three- and one-spacecraft observations, respectively; a similar trend characterizes the errors on the CME speeds/widths (L. Mays, private communication, 2020).

To avoid the uncertainties related to the choice of a particular CME catalog, \citet{ref:DENOLFO2019} used the latitude of the parent flares. To assess the effect of such approximation, we recalculated the $N_{SEP}$ values by using the flare locations as well as the CME directions from DONKI and \citet{ref:GOPALSWAMY2014}. 
In addition, the connection angles $\delta$ (Equation \ref{eq:great-circle}) relative to each spacecraft location were evaluated by using the corresponding Parker-spiral magnetic field line footpoints at 2.5 R$_{\odot}$, which is similar to the typical particle release height inferred for GLE events assuming that particles are accelerated at CME-driven shocks \citep{ref:REAMES2009,ref:GOPALSWAMY2013}, while they were computed at 30 R$_{\odot}$ in \citet{ref:DENOLFO2019}. Consequently, the new $N_{SEP}$ values also account for the relatively small differences associated with the footpoint calculation.

The numbers of $>$500 MeV interplanetary protons $N_{SEP}$ were derived by \citet{ref:DENOLFO2019} using the $>80$ MeV proton spatial distributions 
as proxies of the ones, not directly measurable, of $>$500 MeV protons. In fact, the STEREO proton observations are limited to 100 MeV, and an estimate based on the extrapolation of the spectral fits above 500 MeV is unreliable due to the large associated uncertainties. For this reason, \citet{ref:DENOLFO2019} considered the obtained $N_{SEP}$ values as upper limits. However, this assumption is strictly valid only for well-connected SEP events, while the proton numbers could be larger especially for events with poorer connectivity, if characterized by a narrow spatial distribution. 
In order to compute more conservative $N_{SEP}$ upper limits that take into account the different widths of the spatial distributions at higher energies, we developed the following approach.
\begin{itemize}
\item The $>$80 MeV differential energy spectrum at the shock nose was assumed to be described by an \citet{ref:ER1985} functional form: $I_{n}(E) \propto E^{-\alpha_{n}}\exp(-E/E_{0})$. The spectrum of the 2012 May 17 GLE measured by PAMELA, with parameters $\alpha_{n}$ $\sim$ 2.4 and $E_{0}$ $\sim$ 500 MeV \citep{ref:BRUNO2018}, was used as a proxy of the spectrum at the shock nose.
\item The $E>E_{t}$ energy-integrated intensities at the shock nose ($J_{n}^{E_{t}}=\int_{E_{t}}^{\infty} I_{n}(E) dE$) and at a given spacecraft spherical distance $\delta_{E_{t}}$ ($J_{\delta}^{E_{t}}=\int_{E_{t}}^{\infty} I_{\delta}(E) dE$) are related by:     
$J_{\delta}^{E_{t}} = J_{n}^{E_{t}}C_{\delta}^{E_{t}}$, 
with the angular correction factor given by:
\begin{equation}    
C_{\delta}^{E_{t}} = \exp\left[-\frac{1}{2}\left(\frac{\delta_{E_{t}}}{\sigma_{E_{t}}}\right)^{2}\right],
\end{equation}
where the Gaussian standard deviation $\sigma_{E_{t}}$ describes the width of the corresponding SEP spatial distribution. 
\item The connection angle $\delta_{E_{t}}$ is energy dependent: in fact, the peak of the spatial distribution, on average located on field lines with footpoints westward of the source coordinates, tends to move eastward with increasing energy \citep{ref:BRUNORICHARDSON2021}. 
The SEP peak longitudinal displacement can be parameterized as: 
\begin{equation}\label{eq:longitudinal_shift}    
\phi_{sep}(E) - \phi_{s} = \phi_{0} - \phi_{1} log(E),
\end{equation}
where $\phi_{s}$ is the longitude of the parent flare/CME. 
For this analysis, we set $\phi_{1}$ to 4.81, i.e. the mean value obtained by \citet{ref:BRUNORICHARDSON2021} using a sample of 32 SEP events, while $\phi_{0}$ was computed event by event based on the longitudinal deviation obtained for the $>$80 MeV spatial distribution.
\item The $>$80 MeV and $>$500 MeV energy-integrated intensities at the spherical distance $\delta$ associated with the Earth footpoint, $J_{\delta}^{80}$ and $J_{\delta}^{500}$, were obtained from \citet{ref:DENOLFO2019}, while the ratio of the corresponding energy-integrated spectra at the shock nose, $R_{J}$ = $J_{n}^{500}$ / $J_{n}^{80}$, was derived based on the $\alpha_{n}$ and $E_{0}$ parameters.
\item Then, the $>$500 MeV energy-integrated spectrum at the shock nose was calculated as:
\begin{equation}    
J_{n}^{500} = J_{\delta}^{500} / C_{\delta}^{500} 
= R_{J} \hspace{0.05cm} J_{n}^{80} 
= R_{J} \hspace{0.05cm} J_{\delta}^{80} / C_{\delta}^{80},
\end{equation}
where $C_{\delta}^{80}$ is the angular correction factor based on the $\sigma_{80}$ value computed by \citet{ref:DENOLFO2019} for each of the fourteen SEP events.
\item The corresponding standard deviation was evaluated as:
\begin{equation}
\sigma_{500} = \sqrt{ - \frac{1}{2}\frac{(\delta_{500})^{2}}{log(C_{\delta}^{500})} }.
\end{equation}
with $\delta_{500}$ derived from Equation \ref{eq:longitudinal_shift} -- implying an $\sim$8.8$^{\circ}$ average displacement of the distribution peak from 80 to 500 MeV --
and this value was used in place of $\sigma_{80}$ in the calculation by \citet{ref:DENOLFO2019} to estimate the numbers of interplanetary protons above 500 MeV.
\end{itemize}
We note that all the used quantities -- in particular, the individual $\sigma_{80}$ values -- are based on experimental data. The only assumptions relate to the $\phi_{1}$ value and the shock nose spectrum which, however, was approximated by the hardest SEP spectrum with $>$500 MeV proton data available observed by PAMELA in solar cycle 24. 
The calculated $\sigma_{500}$ standard deviations  
range from $\sim$17\% to $\sim$92\%
of the corresponding $\sigma_{80}$ values.

Special attention was given to those SEP events with no appreciable intensities or with a significant background from a previous eruption at a given spacecraft location, in order to derive conservatively-high upper limits on the corresponding $N_{SEP}$ numbers.
In particular, for the 2011 August 9 and 2012 July 7 events, with no relevant SEP signal at STEREO-B, we assessed the $>$80 MeV spatial distribution based on near-Earth and STEREO-A measurements by evaluating the peak longitude
with Equation \ref{eq:longitudinal_shift}, using both the mean $\phi_{0}/\phi_{1}$ values from \citet{ref:BRUNORICHARDSON2021}, to constrain the fit. The same method was applied to the longitudinally well-connected event on 2011 June 7, characterized by a large background at both STEREOs, for which we  assumed a 20$^{\circ}$ standard deviation at 80 MeV, in contrast to the $\sim$40$^{\circ}$ value used by \citet{ref:DENOLFO2019}.
Finally, for the STEREO-B observations relative to the
events on 2011 March 7, 2012 January 27 and 2013 October 10, significantly influenced by ongoing events, 
we conservatively assumed that the effective, background-subtracted intensities were a factor three lower than assumed by \citet{ref:DENOLFO2019}.

%
%
%
%

\begin{figure}
\centering
\includegraphics[width=\linewidth]{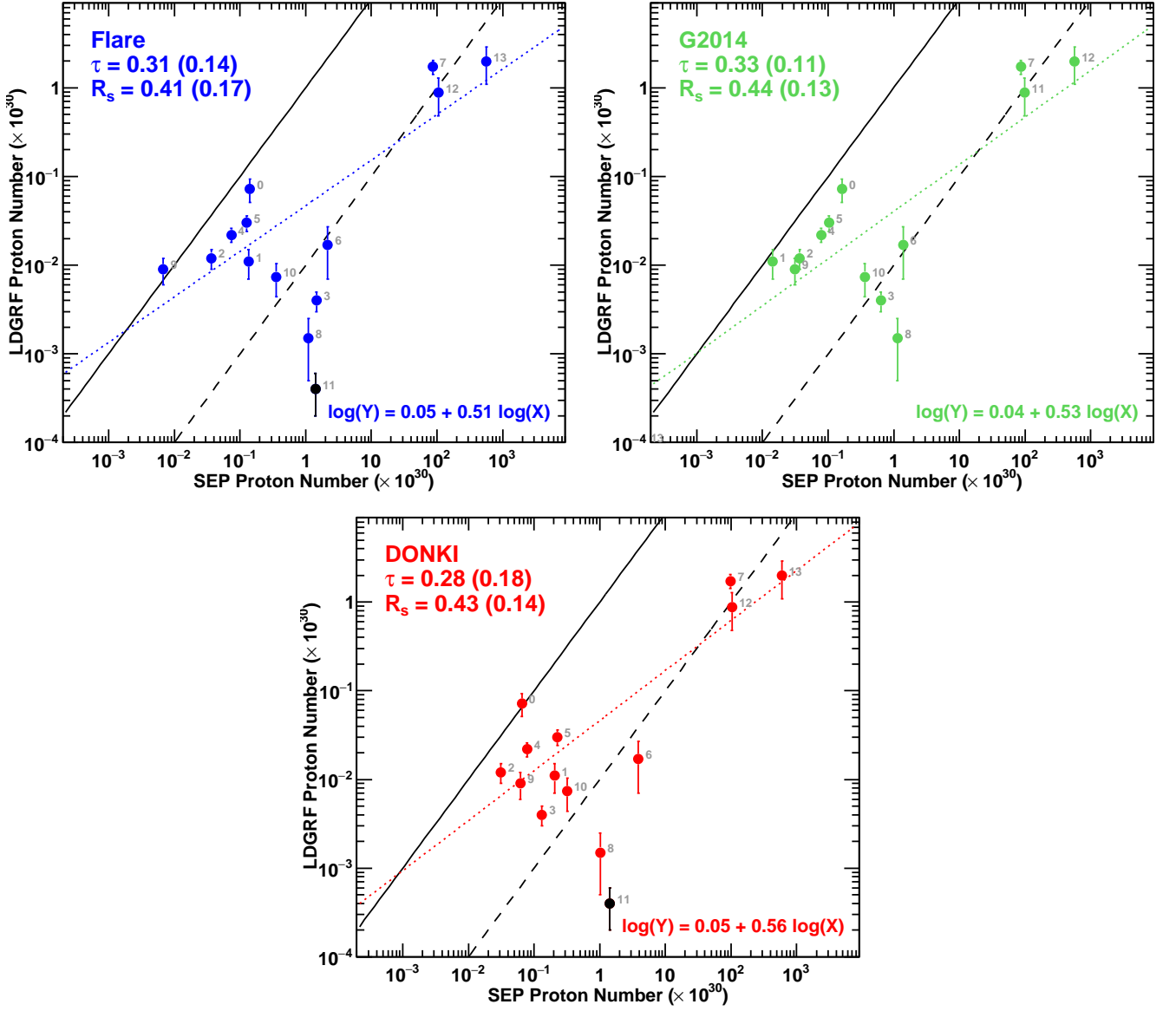}
\caption{Scatter plots of the $N_{LDGRF}$ vs $N_{SEP}$ values obtained by using the upper limits on the numbers of $>$500 MeV interplanetary protons. The three panels report the calculations made implementing the flare latitudes (blue, top left panel) and the CME latitudinal angles from \citet[G2014]{ref:GOPALSWAMY2014} (green, top right panel) and DONKI (red, bottom panel) -- see Table \ref{tab:Table1}, with corresponding Kendall ($\tau$) and Spearman ($R_{s}$) correlation factors and relative p-values between brackets.
The event numbers in Table \ref{tab:Table1} are printed adjacent to the data points.
The colored dotted lines indicates the corresponding bi-logarithmic regression lines, with parameters reported in the bottom-right corners.
The solid and dashed black lines mark the 1-to-1 and the 1-to-100 correspondences, respectively. The black point marks the 2013 October 28 event, not included in the correlation and the regression estimates.}
\label{fig:Figure1}
\end{figure}

\subsection{Results}
The calculated $N_{LDGRF}$ and $N_{SEP}$ numbers are summarized in Table \ref{tab:Table2}, while the corresponding scatter plots are shown in Figure \ref{fig:Figure1}; 
the three panels correspond to the upper limits on the $N_{SEP}$ numbers computed by using the flare locations (blue), and the CME directions from \citet{ref:GOPALSWAMY2014} (green) and DONKI (red) reported in Table \ref{tab:Table1}. 
As discussed in Section \ref{Interacting Protons}, the used $N_{LDGRF}$ values do not account for the heliocentric angle-dependent correction factor for a downward isotropic distribution, in contrast to \citet{ref:DENOLFO2019}.
The corresponding Kendall ($\tau$) and Spearman ($R_{s}$) rank correlation coefficients are reported in each panel along with relative p-values. Similar to \citet{ref:DENOLFO2019}, we did not consider the Pearson coefficient since it is 
greatly 
affected by outliers and the distribution points are not uniformly distributed. 
The correlation is in no case
statistically significant, as suggested by the corresponding p-values, which are higher than the typical threshold of 0.05 for the rejection of the null hypothesis (no correlation). 
However, for the case of the \citet{ref:GOPALSWAMY2014} latitudes, the associated probability is 11\% and 14\%, respectively for the Kendall and the Spearman coefficients, suggesting a certain trend toward significance although the results cannot be considered conclusive. 
In general, the correlation appears to be essentially controlled by the three points in the top-right corner of the plots (events on 2012 March 7, 2014 February 25 and 2014 September 1), since the rest of the sample does not seem to follow any specific trend ($\tau$=-0.07, p=0.79; $R_{s}$=-0.14, p=0.70). We also note that the results do not include the outlying point (black circle) for 
the 2013 October 28 event, since the origin of the associated $\gamma$-ray emission is uncertain \citep{ref:SHARE2018}; in addition, this event is not included in the \citet{ref:GOPALSWAMY2014} calculation (see Table \ref{tab:Table1}). Finally, the corresponding bi-logarithmic regression lines are also shown in Figure \ref{fig:Figure1}, emphasizing that the average $N_{LDGRF}$/$N_{SEP}$ ratio is not constant, 
but decreases with increasing $N_{SEP}$ values.

\begin{figure}
\centering
\includegraphics[width=0.8\linewidth]{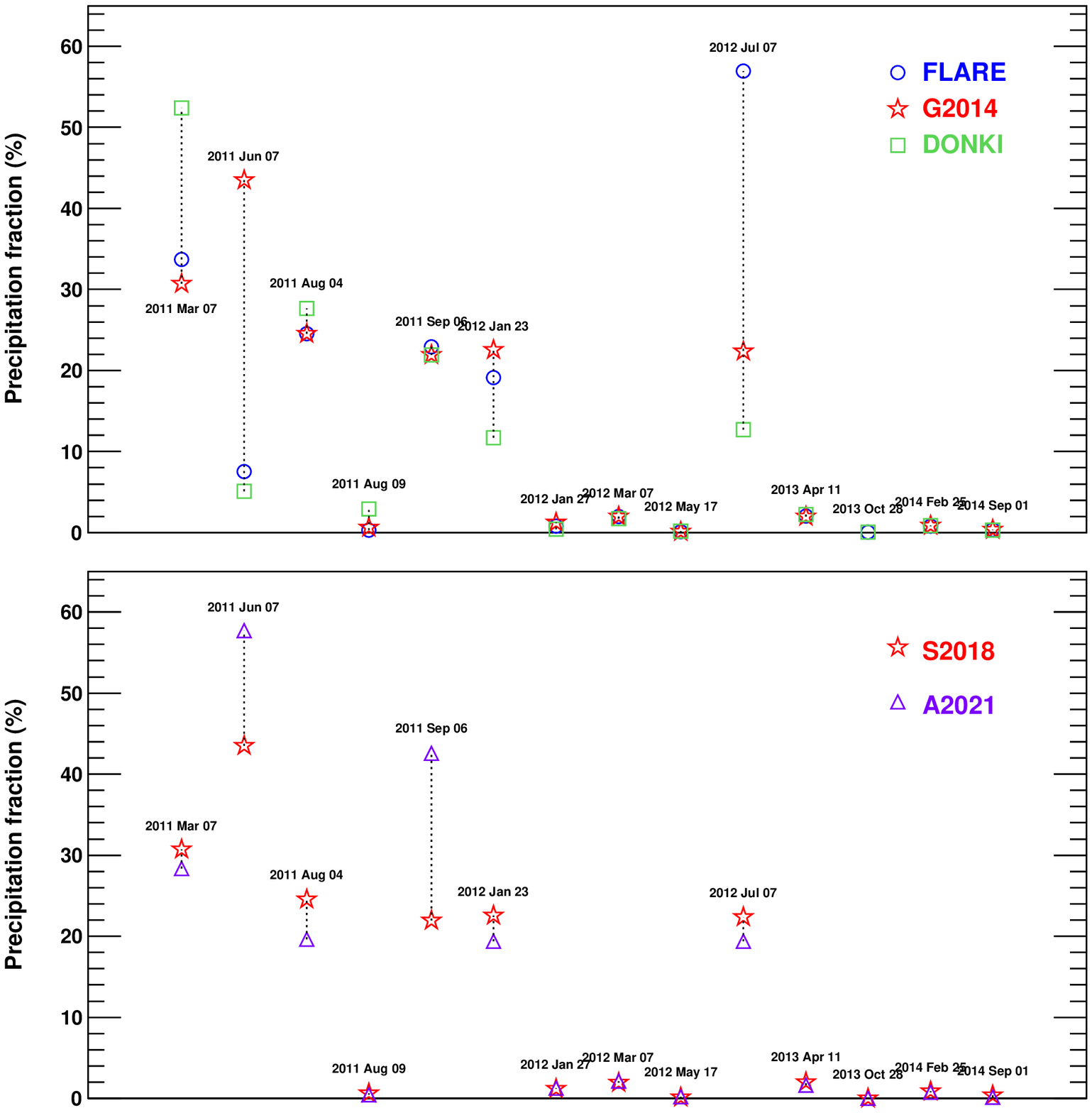}
\caption{Top: lower limits on the precipitation fractions $P_{N}$ = $N_{LDGRF}$ / ($N_{LDGRF}$ + $N_{SEP}$) relative to the proton numbers assuming different latitudinal angles displayed in Figure \ref{fig:Figure1}. Bottom: comparison between the precipitation fraction lower limits using the $N_{LDGRF}$ values from \citet{ref:SHARE2018} (red) and \citet[A2021]{ref:AJELLO2021} (green), and the upper limits on the $N_{SEP}$ numbers derived using the \citet[G2014]{ref:GOPALSWAMY2014} latitudes. The dotted lines are to guide the eye.} 
\label{fig:Figure2}
\end{figure}

The corresponding lower limits on the precipitation fractions are displayed in the top panel of Figure \ref{fig:Figure2} and reported in Table \ref{tab:Table2}.
We note that, despite the conservative assumptions described in the previous Sections, the precipitation fraction is still higher than 10--20\% for five or six of the fourteen analyzed events, depending on the used latitudes. 
Interestingly, no particular trend with CME- or flare-related parameters was found for the $P_{N}$ distribution; in particular low/high-fraction occurrences appear to be independent on the CME speed. On the other hand,
it has been shown by \citet{ref:HUTCHINSON2022} that smaller precipitation fractions tend to occur for longer-duration events associated with faster CMEs.
In contrast, 
a relatively high $P_{N}$ value was obtained for the 2011 March 7 and the 2012 January 23 events, both accompanied by long-duration $\gamma$-ray emission and fast CMEs (2223 km s$^{-1}$ and 2511 km s$^{-1}$), while a low precipitation fraction was found for the short-duration event on 2013 April 11, linked to a 1369 km s$^{-1}$ CME.

\begin{table}[!t]
\center
\scriptsize
\setlength{\tabcolsep}{2.5pt}
\begin{tabular}{l|c|cc|ccc|ccc|ccc|}
& SEP & \multicolumn{2}{c|}{$N_{LDGRF}$} & \multicolumn{3}{c|}{Flare} & \multicolumn{3}{c|}{G2014} & \multicolumn{3}{c|}{DONKI} \\
No & event & S2018 & A2021 & $N_{SEP}$ & $P_{N}^{S2018}$ & $P_{N}^{A2021}$ & $N_{SEP}$ & $P_{N}^{S2018}$ & $P_{N}^{A2021}$ & $N_{SEP}$ & $P_{N}^{S2018}$ & $P_{N}^{A2021}$  \\
\hline
1 & 2011/03/07 & 7.20$\times10^{28}$ & 6.44$\times10^{28}$ & 1.42$\times10^{29}$ & 33.67 & 31.23 & 1.63$\times10^{29}$ & 30.70 & 28.38 & 6.54$\times10^{28}$ & 52.42 & 49.63 \\ 
2 & 2011/06/07 & 1.10$\times10^{28}$ & 1.95$\times10^{28}$ & 1.35$\times10^{29}$ & 7.55 & 12.64 & 1.43$\times10^{28}$ & 43.47 & 57.69 & 2.05$\times10^{29}$ & 5.09 & 8.68 \\ 
3 & 2011/08/04 & 1.20$\times10^{28}$ & 9.00$\times10^{27}$ & 3.69$\times10^{28}$ & 24.56 & 19.63 & 3.69$\times10^{28}$ & 24.56 & 19.63 & 3.13$\times10^{28}$ & 27.69 & 22.31 \\ 
4 & 2011/08/09 & 4.00$\times10^{27}$ & 2.70$\times10^{27}$ & 1.47$\times10^{30}$ & 0.27 & 0.18 & 6.36$\times10^{29}$ & 0.62 & 0.42 & 1.32$\times10^{29}$ & 2.94 & 2.01 \\ 
5 & 2011/09/06 & 2.20$\times10^{28}$ & 5.80$\times10^{28}$ & 7.39$\times10^{28}$ & 22.95 & 43.99 & 7.83$\times10^{28}$ & 21.94 & 42.56 & 7.83$\times10^{28}$ & 21.94 & 42.56 \\ 
6 & 2012/01/23 & 3.00$\times10^{28}$ & 2.47$\times10^{28}$ & 1.27$\times10^{29}$ & 19.11 & 16.28 & 1.03$\times10^{29}$ & 22.57 & 19.35 & 2.27$\times10^{29}$ & 11.69 & 9.83 \\ 
7 & 2012/01/27 & 1.70$\times10^{28}$ & 1.72$\times10^{28}$ & 2.17$\times10^{30}$ & 0.78 & 0.79 & 1.39$\times10^{30}$ & 1.21 & 1.22 & 3.87$\times10^{30}$ & 0.44 & 0.44 \\ 
8 & 2012/03/07 & 1.73$\times10^{30}$ & 1.86$\times10^{30}$ & 8.67$\times10^{31}$ & 1.96 & 2.10 & 8.67$\times10^{31}$ & 1.96 & 2.10 & 9.84$\times10^{31}$ & 1.73 & 1.85 \\ 
9 & 2012/05/17 & 1.50$\times10^{27}$ & 2.29$\times10^{27}$ & 1.10$\times10^{30}$ & 0.14 & 0.21 & 1.14$\times10^{30}$ & 0.13 & 0.20 & 1.02$\times10^{30}$ & 0.15 & 0.22 \\ 
10 & 2012/07/07 & 9.00$\times10^{27}$ & 7.50$\times10^{27}$ & 6.80$\times10^{27}$ & 56.95 & 52.44 & 3.12$\times10^{28}$ & 22.38 & 19.38 & 6.18$\times10^{28}$ & 12.71 & 10.82 \\ 
11 & 2013/04/11 & 7.40$\times10^{27}$ & 6.00$\times10^{27}$ & 3.58$\times10^{29}$ & 2.03 & 1.65 & 3.63$\times10^{29}$ & 2.00 & 1.62 & 3.21$\times10^{29}$ & 2.25 & 1.84 \\ 
12 & 2013/10/28 & 4.00$\times10^{26}$ & \nodata & 1.41$\times10^{30}$ & 0.03 & \nodata & 1.37$\times10^{30}$ & 0.03 & \nodata & 1.43$\times10^{30}$ & 0.03 & \nodata \\ 
13 & 2014/02/25 & 8.80$\times10^{29}$ & 7.19$\times10^{29}$ & 1.05$\times10^{32}$ & 0.83 & 0.68 & 9.83$\times10^{31}$ & 0.89 & 0.73 & 1.03$\times10^{32}$ & 0.84 & 0.69 \\ 
14 & 2014/09/01 & 1.99$\times10^{30}$ & 7.40$\times10^{29}$ & 5.68$\times10^{32}$ & 0.35 & 0.13 & 5.66$\times10^{32}$ & 0.35 & 0.13 & 6.00$\times10^{32}$ & 0.33 & 0.12 \\ 
\end{tabular}
\caption{Summary of estimated proton numbers. The first two columns report the SEP event number and date; columns 3--4 list the
numbers of $>$500 MeV interacting protons ($N_{LDGRF}$) according to \citet[S2018]{ref:SHARE2018} and \citet[A2021]{ref:AJELLO2021}; columns 5--7, 8--10 and 11--13 report the numbers of $>$500 MeV interplanetary protons ($N_{SEP}$) and the corresponding precipitation fractions ($P_{N}$, \%) using the calculations of $N_{SEP}$ using latitudes of the associated flare, or CME latitudes from \citet[G2014]{ref:GOPALSWAMY2014} and the DONKI catalog (see Table \ref{tab:Table1}).}
\label{tab:Table2}
\end{table}

As mentioned in the Section \ref{Interacting Protons}, we compared the $N_{LDGRF}$ numbers from \citet{ref:SHARE2018} to the ones derived by \citet{ref:AJELLO2021}, and found a reasonable agreement for most events. Results in terms of precipitation fractions are displayed in the bottom panel of Figure \ref{fig:Figure2}, where for the upper limits on the $N_{SEP}$ numbers
we used the estimates made with the latitudes from \citet{ref:GOPALSWAMY2014}.
These were assumed to be the most reliable since the CME direction is a likely better approximation of the SEP propagation axis, and the CME parameters in the DONKI catalog are often based on real-time calculations. As aforementioned, the largest variations involve the eruptions on 2011 Jun 7 and 2011 September 6, for which the $N_{LDGRF}$ from \citet{ref:AJELLO2021} are a factor $\sim$1.8 and $\sim$2.6 higher, respectively. 
We note that, for both calculations, $P_{N}$ is greater than 20\% for six events. 

\begin{figure}
\centering
\includegraphics[width=0.8\linewidth]{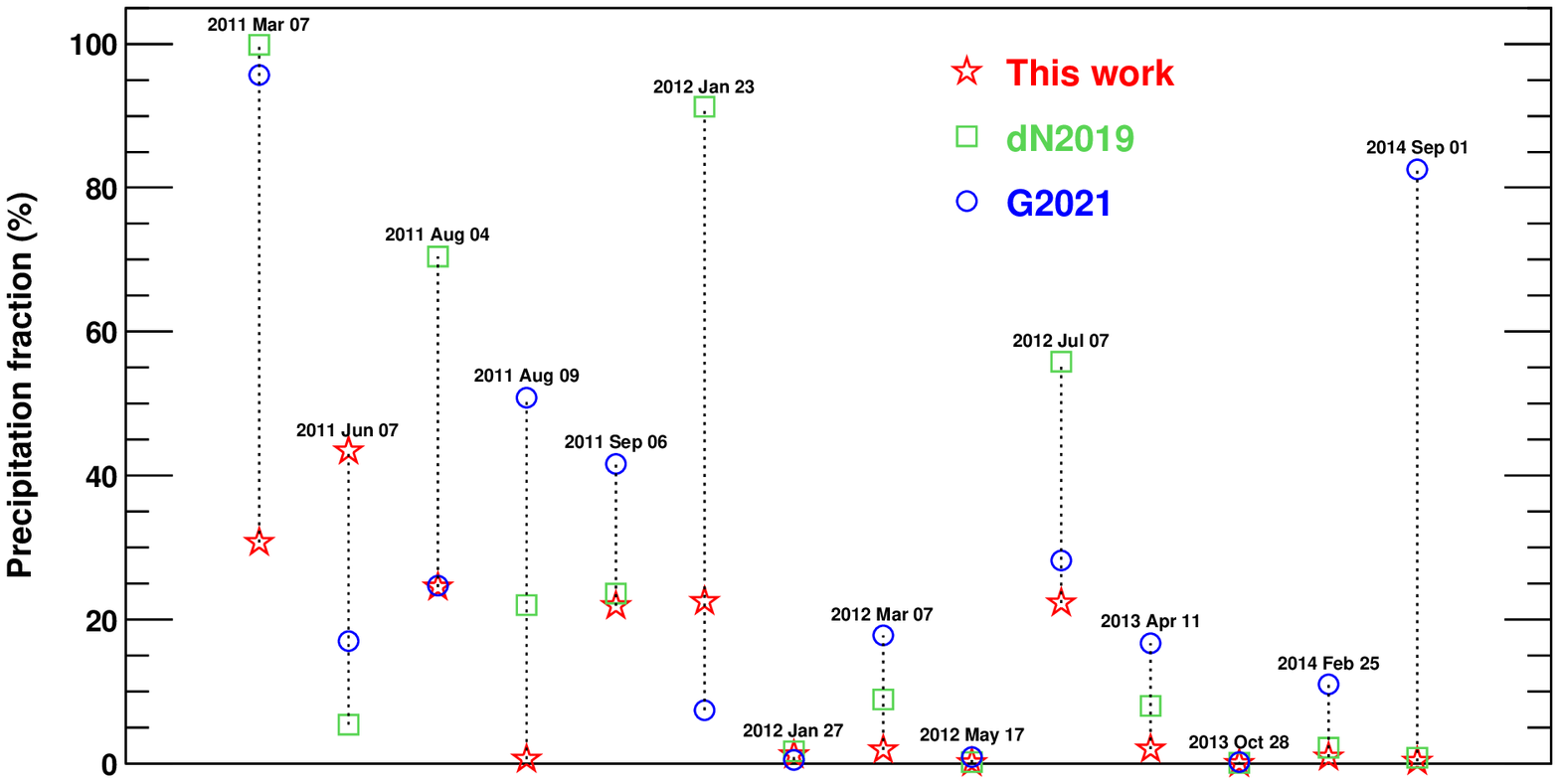}
\caption{Comparison between the precipitation fraction lower limits derived in this work, using the $N_{SEP}$ upper limits based on the CME latitudes from \citet{ref:GOPALSWAMY2014}, and the precipitation fractions estimated by \citet[dN2019]{ref:DENOLFO2019} and \citet[G2021]{ref:GOPALSWAMY2021}. All the calculations rely on the $N_{LDGRF}$ values from \citet{ref:SHARE2018}. The dotted lines are to guide the eye.}
\label{fig:Figure3}
\end{figure}

%
%
%
%

\subsection{Comparison with Other Calculations}\label{Comparison with Other Calculations}
The comparison with the previous work by \citet{ref:DENOLFO2019} is shown in Figure \ref{fig:Figure3}.
It can be noted that the precipitation fractions we obtained are often significantly lower, as a consequence of the removal of the heliocentric angle-dependent correction factor on the $N_{LDGRF}$ numbers, as well as the improved estimate of the SEP distributions above 500 MeV described in the previous Section. However, even accounting for these more conservative assumptions, we confirm the conclusions of \citet{ref:DENOLFO2019} 
about the lack of a statistically-significant correlation between the number of SEP protons and the number of protons required to account for the $\gamma$-ray emission. Furthermore, a high precipitation fraction is required for several of the events that appears inconsistent with a CME-shock origin of LDGRFs.
Figure \ref{fig:Figure3}
also includes results
from \citet{ref:GOPALSWAMY2021}, who modified the proton numbers from \citet{ref:DENOLFO2019} with additional correction factors, resulting in a high statistical correlation between $N_{LDGRF}$ and $N_{SEP}$.
Specifically, they increased the $N_{LDGRF}$ values for eruptions near and behind the limb, assuming an underestimate for the atmospheric attenuation of the $\gamma$-ray flux. In particular, they computed a correction factor of 560 for the 2014 September 1 event. Since this event occurred behind the limb, it cannot be excluded that the flux computed by \citet{ref:SHARE2018} (and \citet{ref:AJELLO2021}) was somewhat underestimated. On the other hand, as LDGRFs are highly extended, 
we would not expect the reconstructed $\gamma$-ray flux to account for only $\sim$0.18\% (i.e. 1/560) of the total emitted flux. 
This large correction factor also increases the precipitation factor by approximately two orders of magnitude, to $\sim$82\%, which cannot be reconciled with the predictions of current models of particle precipitation from a CME shock \citep{ref:HUTCHINSON2022}.   
Furthermore, we note that the $\gamma$-ray occultation was already accounted for by \citet{ref:SHARE2018}, so the final $N_{LDGRF}$ numbers of \citet{ref:GOPALSWAMY2021} are actually based on an over correction for this effect.

Analogously, \citet{ref:GOPALSWAMY2021} over-counted the SEP latitudinal correction factor already implemented by \citet{ref:DENOLFO2019} when calculating $N_{SEP}$. In addition, their correction method relies on some questionable assumptions. In particular, for the shock nose, they assumed a simple inverse power-law energy spectrum with 
index $\alpha_{n}$ = 2 extending to infinite energies. 
However, this is inconsistent with historical neutron monitor observations that show a $\ge$3 index even during the initial phase of major GLEs (e.g., \citealp{ref:VASHENYUK2011,ref:MISHEV2012,ref:KOLDOBSKIY2022,ref:MISHEV2023}). Indeed, their spectral fit studies are based on the analysis of the GOES proton intensities at low energies (10--100 MeV); furthermore, 
the large uncertainties associated with the GOES uncalibrated energy channels typically result in systematically harder spectra \citep{ref:BRUNO2017}. We also note that their energy-dependent width calculation was applied only to five eruptions with high latitudinal angle. On the other hand, given the spherical symmetry assumed for the SEP spatial distribution, the decrease of the associated standard deviation with increasing energy affects both the SEP longitudinal and latitudinal extent.
In fact, we found that correcting the whole event sample with the same criterion results in a systematic shift of the $N_{SEP}$ numbers towards higher values (including the events with largest $N_{SEP}$/$N_{LDGRF}$ ratios), and in a remarkably lower correlation with the $N_{LDGRF}$ numbers. 
Finally, \citet{ref:GOPALSWAMY2021} approximated the CME latitudes with the corresponding plane-of-the-sky position angles from the CDAW catalog, affected by projection effects resulting in systematic errors in the CME direction. For instance, the position angles evaluated for the 2012 January 23 and the 2012 March 7 CMEs are N56 and N33, respectively, compared to the N30 and the N18 CME latitudes computed by \citet{ref:GOPALSWAMY2014} with multi-spacecraft observations using the GCS method.
Thus, for these reasons, we question the validity of the high correlation between the ``corrected'' values of $N_{SEP}$ and $N_{LDGRF}$ claimed by \citet{ref:GOPALSWAMY2021}.

%
%
%
%

\subsection{Interplanetary transport effects}
\begin{figure}
\centering
\includegraphics[width=0.8\linewidth]{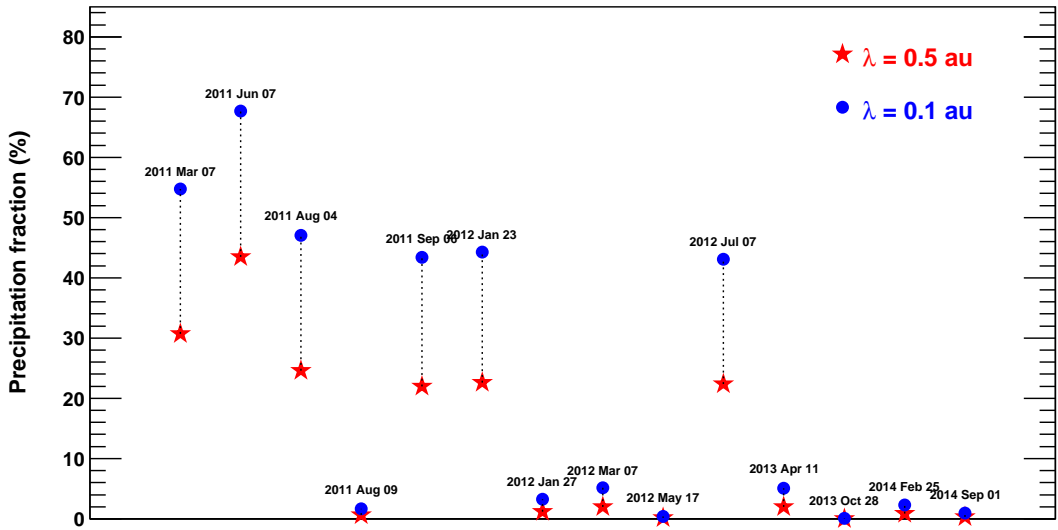}
\caption{Comparison between the precipitation fraction lower limits using the different interplanetary turbulence conditions, as parameterized by the mean free path $\lambda$. Both calculations are based on the $N_{SEP}$ upper limits obtained with the CME latitudes from \citet{ref:GOPALSWAMY2014} and the $N_{LDGRF}$ values from \citet{ref:SHARE2018}.}
\label{fig:Figure4}
\end{figure}

\citet{ref:DENOLFO2019} accounted for some of the interplanetary transport effects on high-energy protons, in particular the multiple crossings experienced by particles at 1 AU due to local scattering that increase the apparent SEP intensity, by means of 3D test particle simulations including the effects of the HCS. In particular,
they conservatively assumed a mean free path $\lambda$ = 0.5 au. 
This value corresponds to an average number of 1 AU crossings of 8 and 11 for periods of solar magnetic polarity $A^{+}$ and $A^{-}$, respectively (see their Table 4).
Consequently, the calculated $N_{SEP}$ values -- corrected by the corresponding number of 1 AU crossings -- are expected to be overestimated, and the $P_{N}$ value therefore underestimated, under more turbulent conditions.
We note that a smaller mean free path of 0.3 AU was evaluated for the 2012 May 17 event using PAMELA measurements \citep{ref:DENOLFO2019,ref:DALLA2020}.
Considering an even smaller mean free path of 0.1 AU, the numbers of 1 AU crossings computed by \citet{ref:DENOLFO2019} are 21 and 30, respectively for $A^{+}$ and $A^{-}$ polarity, so the resulting $N_{SEP}$ numbers would be 62--63\% smaller. While these estimates do not affect the level of correlation with the $N_{LDGRF}$ numbers since the intensity is changed by a similar factor for all events, they have noteworthy influences on the corresponding precipitation fractions.  Figure \ref{fig:Figure4} demonstrates the effect of using a higher level of scattering in the estimate of the number of crossings. It can be seen that the fractions are considerably higher for $\lambda$ = 0.1 AU, with a $\sim$40\% minimum value for six events. We note that these values would be even higher using the $N_{LDGRF}$ values from \citet{ref:AJELLO2021}, or the flare/DONKI latitudes in the $N_{SEP}$ calculation.

There are also uncertainties in modeling the SEP spatial distribution,
assumed to be described by a 2D Gaussian given the limited number of observation points (typically three or less). It is likely that the ``true'' distribution may substantially deviate from a Gaussian and, in particular, it may include longitudinal and latitudinal asymmetries, for example related to solar-wind structures (e.g., \citealp{ref:LARIO2022}). In addition, the propagation of high-energy SEPs has been shown to be strongly influenced by the heliospheric current sheet (HCS) and drifts associated with the gradient and curvature of the Parker spiral magnetic field, with significant differences between epochs of $A^{+}$ and $A^{-}$ polarities \citep{ref:DALLA2020}. Furthermore, SEP events with a source region closer
to the HCS ($<$10$^{\circ}$) were found to be more likely associated with GLE events \citep{ref:WATERFALL2022}.

%
%
%
%
\begin{table}[!t]
\setlength\extrarowheight{-0.5pt}
\setlength{\tabcolsep}{12pt}
\center
\begin{tabular}{c|c|cc|cc|cc|c}
\small
 & & \multicolumn{2}{c|}{G2019} & \multicolumn{2}{c|}{S2018} & \multicolumn{2}{c|}{W2018} & A2021 \\
\# & Date & $T_{start}$ & $\Delta T$ & $T_{start}$ & $\Delta T$ & $T_{start}$ & $\Delta T$ & $\Delta T$ \\
\hline
1 & 2011/03/07 & 20:12 & 21.02 & 20:00 & 15.00 & 20:15 & 10.10 & 15.80 \\ 
2 & 2011/06/02 & 07:46 & 6.98 & 08:10 & 3.83 & 09:43 & 0.70 & \nodata \\ 
3 & 2011/06/07 & 06:41 & 3.08 & 07:00 & 3.00 & 07:34 & 0.90 & 6.00 \\ 
4 & 2011/08/04 & \nodata & \nodata & 04:10 & 3.00 & 04:59 & 0.60 & 2.30 \\ 
5 & 2011/08/09 & \nodata & \nodata & 08:03 & 0.05 & \nodata & \nodata & 0.87 \\ 
6 & 2011/09/06 & \nodata & \nodata & 22:21 & 0.98 & 22:13 & 0.60 & 2.00 \\ 
7 & 2011/09/07 & \nodata & \nodata & 22:45 & 2.42 & 23:36 & 1.00 & 2.02 \\ 
8 & 2011/09/24 & \nodata & \nodata & 09:40 & 0.07 & \nodata & \nodata & 1.20 \\ 
9 & 2012/01/23 & 03:59 & 15.43 & 04:20 & 7.67 & 04:07 & 5.20 & 5.90 \\ 
10 & 2012/01/27 & 18:37 & 3.59 & 19:00 & 3.00 & 19:44 & 1.90 & 6.80 \\ 
11 & 2012/03/05 & 04:09 & 4.25 & 04:30 & 5.50 & 04:12 & 3.60 & 4.40 \\ 
12 & 2012/03/07 & 00:24 & 20.47 & 00:28 & 19.55 & 00:45 & 19.50 & 20.30 \\ 
13 & 2012/03/09 & 03:53 & 8.85 & 04:30 & 6.00 & 05:16 & 3.80 & 7.20 \\ 
14 & 2012/03/10 & 17:44 & 11.62 & 20:00 & 6.00 & \nodata & \nodata & 6.00 \\ 
15 & 2012/05/17 & 01:47 & 3.08 & 02:10 & 3.17 & 02:17 & 1.90 & 2.60 \\ 
16 & 2012/06/03 & \nodata & \nodata & 17:54 & 1.10 & 17:39 & 0.40 & 1.90 \\ 
17 & 2012/07/06 & \nodata & \nodata & 23:14 & 1.77 & 23:18 & 0.90 & 1.27 \\ 
18 & 2012/10/23 & \nodata & \nodata & 03:20 & 2.00 & \nodata & \nodata & 1.90 \\ 
19 & 2012/11/27 & \nodata & \nodata & 15:55 & 0.27 & 15:49 & 0.80 & 0.17 \\ 
20 & 2013/04/11 & \nodata & \nodata & 07:10 & 0.33 & 07:01 & 0.60 & 0.38 \\ 
21 & 2013/05/13 & 02:17 & 7.47 & 02:30 & 3.50 & 04:32 & 0.70 & 4.00 \\ 
22 & 2013/05/13 & 16:05 & 8.69 & 17:00 & 6.00 & 17:17 & 3.80 & 6.10 \\ 
23 & 2013/05/14 & 01:11 & 5.99 & 01:20 & 5.67 & 01:12 & 5.40 & 5.90 \\ 
24 & 2013/05/15 & 01:48 & 3.61 & 02:00 & 7.00 & \nodata & \nodata & 3.50 \\ 
25 & 2013/10/11 & \nodata & \nodata & 07:14 & 0.27 & 06:32 & 1.10 & 0.38 \\ 
26 & 2013/10/25 & \nodata & \nodata & 08:02 & 1.47 & 08:20 & 0.60 & 1.40 \\ 
27 & 2013/10/28 & \nodata & \nodata & 15:20 & 1.67 & 15:34 & 1.20 & 1.60 \\ 
28 & 2014/01/06 & \nodata & \nodata & \nodata & \nodata & 07:25 & 1.10 & 0.27 \\ 
29 & 2014/01/07 & \nodata & \nodata & \nodata & \nodata & 18:42 & 0.80 & 1.05 \\ 
30 & 2014/02/25 & 00:49 & 8.46 & 00:50 & 4.67 & 01:11 & 6.60 & 8.40 \\ 
31 & 2014/09/01 & 11:11 & 3.92 & 11:02 & 4.83 & 11:02 & 1.90 & 2.50 \\ 
32 & 2014/09/10 & \nodata & \nodata & \nodata & \nodata & 17:47 & 0.60 & 0.30 \\ 
33 & 2015/06/21 & 02:36 & 14.06 & 02:20 & 11.67 & 05:24 & 0.50 & 11.50 \\ 
34 & 2015/06/25 & \nodata & \nodata & \nodata & \nodata & 09:25 & 0.70 & 2.40 \\ 
35 & 2017/09/06 & 12:02 & 18.43 & \nodata & \nodata & \nodata & \nodata & 13.33 \\ 
36 & 2017/09/10 & 16:06 & 15.18 & \nodata & \nodata & \nodata & \nodata & 13.90 \\ 
\end{tabular}
\caption{Onset times (hh:mm) and durations (hours) for the LDGRF events analyzed by \citet[G2019]{ref:GOPALSWAMY2019b}, \citet[S2018]{ref:SHARE2018}, \citet[W2018]{ref:WINTER2018} and \citet[A2021]{ref:AJELLO2021}. The dots (...) indicate events not included in the calculation.}
\label{tab:Table3}
\end{table}

\section{LDGRFs and type-II radio emission}\label{Gamma-ray and type-II radio emission}
Possible support for the CME-shock origin of LDGRF events comes from the correlation between LDGRF durations and the ending frequencies and durations of the interplanetary type-II bursts associated with the CMEs accompanying the $\gamma$-ray emission reported by 
\citet{ref:GOPALSWAMY2018}, since this radio emission is indicative of particle acceleration occurring at a shock. More extended shock acceleration, as indicated by the duration of the emission and its ending frequency (lower frequencies correspond to larger distances from the Sun), might be expected to lead to longer-duration particle precipitation and LDGRFs. 
To gain further insight into the interpretation of this result and its significance, we compared the $\gamma$-ray event parameters used by \citet{ref:GOPALSWAMY2018}, with those obtained with three independent estimates by \citet{ref:SHARE2018}, \citet{ref:WINTER2018} and \citet{ref:AJELLO2021}. In the case of the dataset by \citet{ref:AJELLO2021}, only the LDGRF durations were provided (see their Table 1). Several important differences can be pointed out. We first note that, unlike \citet{ref:SHARE2018} and \citet{ref:WINTER2018} who analyzed 33 and 29 LDGRFs, respectively, \citet{ref:GOPALSWAMY2018} only considered the sub-sample of the thirteen largest $\gamma$-ray events with durations exceeding $\sim$5 hours. 
The minimum duration requirement was justified by the the need to avoid any association with impulsive-flare emission.
Subsequently,
\citet{ref:GOPALSWAMY2019b} reported that the relationship between $\gamma$-ray emission times and type-II end frequencies and durations also holds after the inclusion of events with $>$3 hour duration.
The temporal information about the interplanetary radio data are based on the Wind/WAVES type-II burst catalog, 
with refined emission ending times. 

\begin{figure}
\centering
\includegraphics[width=\linewidth]{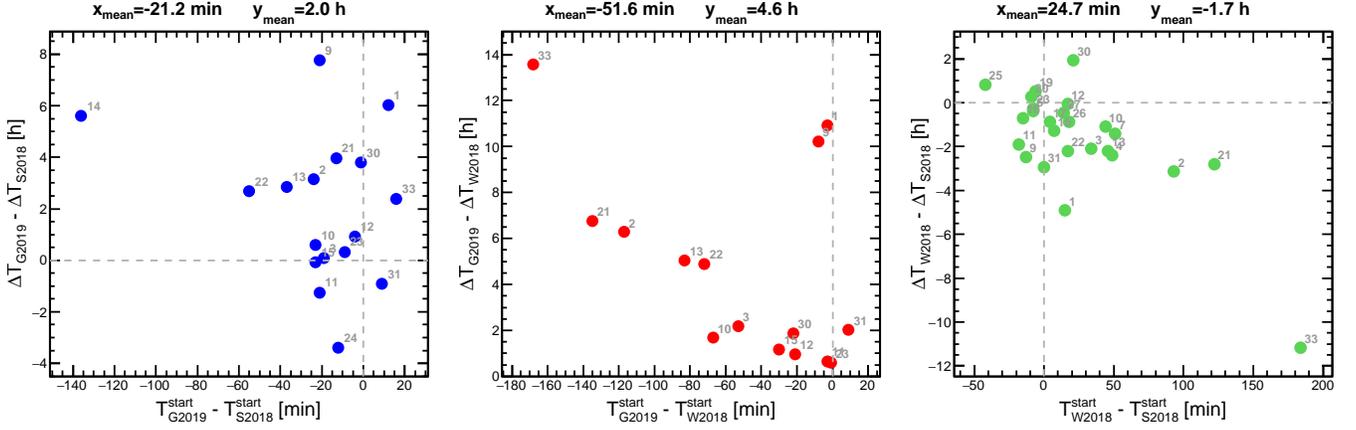}
\caption{Scatter plots of the difference between the $\gamma$-ray event durations $\Delta T$ vs the difference between the corresponding onset times $T^{start}$, based on the estimates from \citet[G2019]{ref:GOPALSWAMY2019b}, \citet[S2018]{ref:SHARE2018} and \citet[W2018]{ref:WINTER2018}. The event numbers in Table \ref{tab:Table3} are printed adjacent to the data points.}
\label{fig:GammaTimesComp}
\end{figure}

Another important difference regards the estimation of the event start $T^{start}$ and stop $T^{stop}$ times with the $Fermi$/LAT data.
\citet{ref:GOPALSWAMY2019b} took $T^{start}$ as the GOES soft X-ray peak time ``to avoid the impulsive phase'', while $T^{stop}$ was obtained as the mid-time between the last data point above the background and the next data point, based on the $Fermi$/LAT ``light-bucket'' time profiles also used by \citet{ref:SHARE2018}. 
However, as reported in Table \ref{tab:Table3} and demonstrated in Figure \ref{fig:GammaTimesComp}, their start and even end times are significantly different from those evaluated by \citet{ref:SHARE2018} who, in particular, fit the onset phase based on higher-resolution (4-minute) LAT data, extrapolating to the background level.
Similar discrepancies are obtained with the values by \citet{ref:WINTER2018}, which are based on the maximum-likelihood method used by the LAT team. The \citet{ref:AJELLO2021} results are not included in Figure \ref{fig:GammaTimesComp} since they did not provide the onset times; however, their event durations are closest to those estimated by \citet{ref:SHARE2018}.
\citet{ref:GOPALSWAMY2019b} estimated $T^{start}$ values that
occur, on average for the sample of events in common,
$\sim$21.2 and $\sim$51.6 minutes earlier with respect to the calculations by \citet{ref:SHARE2018} and \citet{ref:WINTER2018}.
In some cases, the differences are larger than two hours.
Secondly, their $T^{stop}$ values are typically later with respect to those from the same authors. As a result, the event durations calculated by \citet{ref:GOPALSWAMY2019b} are, on average, $\sim$2.0 and $\sim$4.6 hours longer with respect to those computed by \citet{ref:SHARE2018} and \citet{ref:WINTER2018}, respectively. 
The discrepancy is notable for some events.
As example, the durations estimated by \citet{ref:GOPALSWAMY2019b} for the 2011 March 7 and the 2012 January 23 events are $\sim$5--11 hours and $\sim$8--10 hours longer, respectively, bringing them substantially closer to the durations of the corresponding interplanetary type-II bursts. However, as demonstrated by the right panel in Figure \ref{fig:GammaTimesComp}, a significant discrepancy also exists between the \citet{ref:SHARE2018} and \citet{ref:WINTER2018} values, highlighting the large uncertainties characterizing the estimates of the LDGRF times as a result of the \textit{Fermi}-LAT limited time resolution and duty cycle.

An interesting aspect regards the onset time of the 2011 March 7 LDGRF. We note that, even using the most delayed $T_{start}$ value computed by \citet{ref:WINTER2018} (20:15UT), the associated emission commenced $\sim$11 and $\sim$10 
minutes earlier than the solar particle release times\footnote{For a comparison with photon arrival times measured at the observer's distance (1 AU),
derived release times account for an eight minute delay.} computed with the velocity dispersion
analysis by \citet{ref:KLEIN2018} and \citet{ref:XIE2016}, respectively, 
which is inconsistent with a CME-shock origin. Furthermore, the 20:20 UT LDGRF onset time of \citet{ref:SHARE2018} and the later release time derived by other authors (e.g, 20:37UT according to \citealp{ref:PAASSILTA2017}) make the CME association even more unlikely. Similar conclusions were also drawn by \citet{ref:KAHLER2018}.

\begin{figure}
\centering
\includegraphics[width=0.77\linewidth]{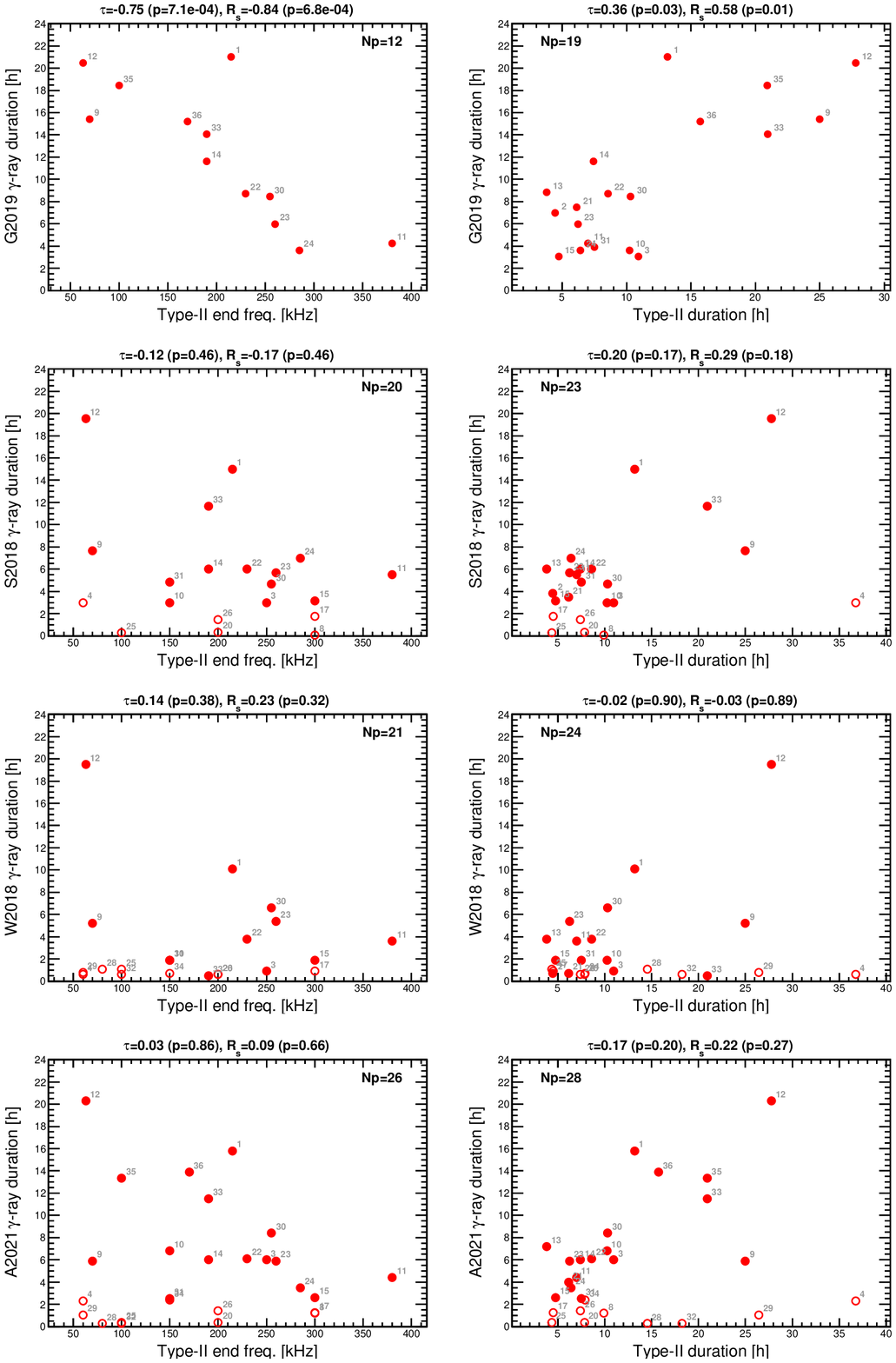}
\caption{Scatter plots of the LDGRF duration as a function of the type-II end frequency (left panels) and duration (right panels). 
From top to bottom: the calculation using results from \citet[G2019]{ref:GOPALSWAMY2019b}, \citet[S2018]{ref:SHARE2018}, \citet[W2018]{ref:WINTER2018} and \citet[A2021]{ref:AJELLO2021}. The number of events used in each plot, along with the Kendall ($\tau$) and Spearman ($R_{s}$) correlation coefficients (and corresponding p-values) are reported for each panel. The event numbers in Table \ref{tab:Table3} are printed adjacent to the data points. The empty markers indicate  events not included by \citet{ref:GOPALSWAMY2019b}.} 
\label{fig:cAll_Comp}
\end{figure}

The implications of using different estimates of LDGRF duration in terms of correlation between the $\gamma$-ray and the radio emissions are demonstrated in Figure \ref{fig:cAll_Comp}. The left and right panels show the scatter plots of the $\gamma$-ray durations vs type-II end frequencies and durations, respectively.
The top panels display the original comparison by \citet{ref:GOPALSWAMY2019b}, while the other three panel rows report the corresponding plots obtained by using the LDGRF durations of \citet{ref:SHARE2018}, \citet{ref:WINTER2018} and \citet{ref:AJELLO2021}. The number of used events (listed in the plots) is larger for the last three samples, accounting for events with durations $<$ 3 hours. In each plot, the radio parameters are taken from \citet{ref:GOPALSWAMY2019b}; for the events not present in \citet{ref:GOPALSWAMY2019b}, we referred to the radio data from the WIND/WAVES type-II burst catalog.
The empty points mark the event sub-sample not included by \citet{ref:GOPALSWAMY2019b}.
As shown by the Kendall and Spearman coefficients at the top of the panels,
the anti-correlation between the LDGRF duration and the type-II ending frequency, and the correlation between the LDGRF and the type-II durations, are significant using the \citet{ref:GOPALSWAMY2019b} data, as reported in that paper. 
However, the level of correlation drastically decreases 
when using the other three estimates of the LDGRF duration, even
using the Pearson coefficient, which is more sensitive to outliers and not used in this analysis.
As discussed above, the durations of a few LDGRF events (e.g., 2012 March 7 and 2012 January 23) were substantially overestimated by \citet{ref:GOPALSWAMY2019b} 
compared to the other studies, and the scatter plots in Figure \ref{fig:cAll_Comp} using the LDGRF durations from \citet{ref:SHARE2018}, \citet{ref:WINTER2018} and \citet{ref:AJELLO2021} have quite different distributions with little evidence of the correlations evident using the \citet{ref:GOPALSWAMY2019b} parameters.
Given the statistical limitations, at most Figure \ref{fig:cAll_Comp} suggests that the longest-duration LDGRFs tend to be associated with longest DH type-II bursts or lowest end frequencies, 
whereas LDGRFs can also have a range of durations for a given type-II duration or end frequency.  
Furthermore, any notable relationship disappears by removing the longest-duration point (2012 March 7), as confirmed by the low values of the rank correlation coefficients. 
It should be noted that the samples analyzed by \citet{ref:SHARE2018} and \citet{ref:WINTER2018} (limited to before 2015) do not include the two long-duration events on 2017 September 6 and 10 considered by \citet{ref:GOPALSWAMY2019b} and \citet{ref:AJELLO2021}; nevertheless, adding these points to the corresponding scatter plot (using the values computed by \citet{ref:GOPALSWAMY2019b} or \citet{ref:AJELLO2021}) does not cause any appreciable improvement in the correlation.
We conclude that 
the strong relationships between type-II radio emissions and LDGRF durations reported by \citet{ref:GOPALSWAMY2018} are not evident using independent estimates of the LDGFR durations from three other studies, and might be caused by the selection of events with $>$3 hour duration along with a less accurate approach in the analysis of LAT data (uncertainties on LDGRF durations and onset times).

%
%
%
%
\section{Discussion and Conclusions}\label{Conclusions}
In this work we have investigated the uncertainties associated with 
the estimates of
the $>$500 MeV populations of interacting protons producing the high-energy $\gamma$-ray emission in LDGRFs and of escaping protons in the interplanetary space associated with SEP events.
This relationship potentially provides a test of whether the back-precipitation of SEP protons accelerated by CME-driven shocks is able to account for the measured $\gamma$-ray emission. 
In particular, in view of their significant implications on the LDGRF origin, we have re-analyzed the calculation by \citet{ref:DENOLFO2019} of the number of high-energy SEPs for a sample of fourteen SEP events observed at three spacecraft locations. We have reviewed the assumptions used to evaluate the proton numbers for these two populations,
thereby testing their conclusion that there is no significant correlation between these numbers, and also confirming that they imply unrealistically high precipitation fractions.  We have also explored the relationship between the durations of LDGRFs and type-II radio emission reported by \citet{ref:GOPALSWAMY2018} using independent assessments of the LDGRF duration. 
The main conclusions can be summarized as follow.
\begin{enumerate}
    
    \item We have examined the uncertainties associated with the different elements of the calculation, 
    and have used conservative assumptions to estimate lower limits on the 
    precipitation fractions. 
    In particular, 
    we have removed the correction introduced by \citet{ref:SHARE2018} that assumed an isotropic distribution for the downward propagating protons producing the LDGRFs. 
    Furthermore, we have developed a new method to compute the $>$500 MeV SEP spatial distribution based on experimental constraints, and we have discussed the effect of the choice of the latitudinal angle describing the SEP propagation axis using both the parent flare locations and the CME directions from different datasets. We have shown that, even in the most favorable case, the $N_{LDGRF}$ and $N_{SEP}$ numbers are only poorly/moderately correlated.
    Most importantly, although the corresponding precipitation fractions are often substantially lower than those computed by \citet{ref:DENOLFO2019}, the values obtained for several events are still excessively high ($>$20\%) to be compatible with current theories of a CME-shock origin of LDGRFs, that predict a maximum $P_{N}$ value of $\sim$2\% \citep{ref:HUDSON2018,ref:KLEIN2018,ref:HUTCHINSON2022}. 
    
    \item For the numbers of $>$500 MeV interacting protons we used the estimates provided by \citet{ref:SHARE2018} and, more recently, \citet{ref:AJELLO2021}. Overall, the two calculations are in reasonable agreement, although $P_{N}$ is higher for a couple of events in the latter case,
    supporting the lack of a highly-significant correlation with the $N_{SEP}$ numbers and the high precipitation fractions that challenge the CME-shock scenario. 
        
    \item We also note that even larger precipitation fractions would be required for more turbulent interplanetary conditions, or if $N_{LDGRF}$ is underestimated when corrections for occultation effects are made for near- or behind-the-limb eruptions \citep{ref:GOPALSWAMY2021}.
        
\item In addition, we have investigated the relationship between the duration of LDGRFs and the duration and the end frequency of the concomitant interplanetary type-II radio emission. We have found that the high correlation levels obtained by \citet{ref:GOPALSWAMY2018,ref:GOPALSWAMY2019b} are reduced when 
    we use the LDGRF durations from independent analyses by \citet{ref:SHARE2018}, \citet{ref:WINTER2018} and \citet{ref:AJELLO2021}, and when events with durations less than 3 hours, excluded by \citet{ref:GOPALSWAMY2018,ref:GOPALSWAMY2019b}, are included.   

\end{enumerate}

Furthermore, we note that the direct link between the time histories of long-duration $\gamma$-ray and interplanetary type-II emissions is not obvious.
In particular, while LDGRFs are mostly produced by $>$300 MeV protons accelerated 
close to the Sun, type-II radio emission is initiated by the acceleration at shocks of low-energy electrons (e.g., \citealp{ref:BALE1999}).  
Since the emission occurs at the plasma frequency, which decreases with heliocentric distance, a lower end frequency implies that type-II emission was produced by the shock out to larger distances from the Sun.  The correlations found by \citet{ref:GOPALSWAMY2018} therefore suggest that the longest duration LDGFRs are associated with CME shocks that produce type-II radio bursts out to larger heliocentric distances.  Since type-II emission is taken as an indicator that the shock is accelerating particles, then the correlation between the type-II and LDGRF emission durations appears to be consistent with the scenario in which the precipitation of shock accelerated protons generates the LDGRF. However, as noted above, the fraction of precipitating
protons in the CME-shock paradigm is predicted to drastically decrease with increasing shock distances due to the magnetic mirroring effect \citep{ref:HUTCHINSON2022}.  
For example, the $\sim$190 kHz ending frequency in the 2015 June 21 event corresponds to $\sim$90 R$_{\odot}$ \citep{ref:GOPALSWAMY2018}
while, in several cases, the type-II emission continues until the shocks reach the observing spacecraft at 1 AU.  Hence, it is not obvious why more extended type-II bursts and a lower end frequency should be correlated with the LDGRF duration. Furthermore, evidence for the continuous acceleration of $>$300 MeV protons by CME-driven shocks moving out through the inner heliosphere is lacking. 
Since faster CME-driven shocks typically accelerate particles over larger helio-distances, the relationship with DH radio bursts could be just a reflection of the association of LDGRFs with relatively fast CMEs. Consequently, the fraction of precipitating protons in the CME-shock paradigm is predicted to drastically decrease with increasing shock distances due to the magnetic mirroring effect, exacerbating the problem for fast CMEs \citep{ref:HUTCHINSON2022}.

In general, back-precipitation is also disfavored by
the presence of the CME structure following the shock (e.g., \citealp{ref:ZURBUCHEN2006}), which potentially complicates the path of particles propagating from the shock to the Sun. In fact, although magnetic field lines in the CME may be rooted at the Sun and might provide a conduit for particles to precipitate back to an extended region around the solar event, there are two issues with this scenario. First, magnetic mirroring must still be overcome, and the generally low magnetic field fluctuation levels in interplanetary CMEs  will likely reduce the level of particle scattering.  Second, field lines passing through the shock do not enter the CME, and observations typically show a significant energy-dependent drop in the particle intensity between the shock and the interior of the CME (e.g., \citealp{ref:SANDERSON2003}).  Thus, we would expect only a fraction of the particles accelerated at the shock to enter the CME.  In addition, the sheath region behind a shock is typically turbulent and is known to inhibit the propagation of galactic cosmic rays in Forbush decreases.  Therefore, the propagation of high energy particles from the shock is likely to be restricted by the sheath.  Thus, we would expect these model calculations to overestimate the number of back-precipitating protons that might be expected in the presence of CME-related structures following the shock.

These analyses of whether LDGRFs may be accounted for by the CME-shock paradigm are based on assessing correlations between physical parameters, but a strong correlation does not necessarily imply a causal connection.  
In fact, while one would expect a significant statistical relationship between LDGRFs and shock-related phenomena in the CME-shock hypothesis, 
even a high correlation between the interacting and interplanetary ion populations would not necessarily demonstrate that they are both of CME-shock origin.
We have shown that even making conservative assumptions when estimating $N_{LDGRF}$ and $N_{SEP}$, the precipitation fractions for some events remain high and are inconsistent with models of proton back precipitation.
This analysis strongly suggests that the computed numbers of high-energy SEPs appear to be greatly insufficient for producing the observed $\gamma$-ray emission. We also note that most energetic SEP events measured at 1 AU may include a direct contribution of flare-accelerated particles, as suggested by several authors (e.g., \citealp{ref:GRECHNEV2008,ref:MCCRACKEN2008,ref:MASSON2009,ref:KAHLER2017,ref:KOCHAROV2020}). If so, 
the estimated $P_{N}$ values -- based on a pure shock acceleration particle origin -- would be further underestimated. 

As also suggested by the reduction in the correlation between the $\gamma$-ray and the interplanetary type-II radio emissions when including events with a duration $<$3 hours, 
the apparent relationship (with clear exceptions) between LDGRFs and the interplanetary radio emission, as well as the CME speed/width and the SEP event size -- as already noted by \citet{ref:SHARE2018} -- might just be a manifestation of the so-called ``Big Flare Syndrome'' \citep{ref:KAHLER1982}, 
i.e., energetic phenomena are statistically more likely to occur together in large solar eruptions even when there is no specific physical process relating them.
For example, the duration of the LDGRF emission is also moderately correlated with the hard and soft X-ray durations of the associated solar flare emissions, although they originate from unrelated processes. Similarly, faster CMEs tend to be associated with more intense flares.

A natural alternative to the CME-shock scenario is represented by the flare-loop model \citep{ref:RYANLEE1991,ref:MANDZHAVIDZERAMATY1992,ref:CHUPPRYAN2009,ref:GRECHNEV2018,ref:RYAN2018,ref:DENOLFO2019,ref:RYAN2019,ref:DENOLFO2021}. If particles are produced and injected on closed field lines, all of them will tend to precipitate on the solar atmosphere after a relatively long residence time associated with particle trapping and second-order Fermi acceleration.
In addition, large ($\gtrsim$1 R$_{\odot}$) coronal loops can easily explain the extended $\gamma$-ray emission observed for behind-the-limb events. 
Thus, if we decouple, as the data suggest, the two energetic particle populations, we are apparently restricted to an acceleration/transport process that must operate close to the Sun for extended periods of time. However, such models have been discussed in the literature for a loop-based acceleration/trapping process that must be further explored.

\section*{Data Availability Statement}
The DONKI catalog (\url{https://ccmc.gsfc.nasa.gov/donki/}) is compiled at the Community Coordinated Modeling Center (CCMC) by NASA.
The CDAW SOHO/LASCO CME catalog (\url{https://cdaw.gsfc.nasa.gov/CME_list/}) is generated and maintained at the CDAW Data Center by NASA and The Catholic University of America in cooperation with the Naval Research Laboratory. 
The catalog of type-II radio bursts observed by Wind/WAVES can be found at \url{https://cdaw.gsfc.nasa.gov/CME_list/radio/waves_type2.html}. The Lockheed-Martin SolarSoft system is available at the LMSAL website (\url{http://www.lmsal.com/solarsoft/}).

\begin{acknowledgments}
A.~B., G.~A.~D. and I.~G.~R. acknowledge support from the NASA programs NNH19ZDA001N-HSR and NNH19ZDA001N-LWS. S.~D. acknowledges support from the UK STFC through grant ST/V000934/1 and NERC through grant NE/V002864/1.
The authors also thank the anonymous reviewer whose comments helped to improve and clarify this manuscript.
\end{acknowledgments}

\bibliography{Bibliography}{}
\bibliographystyle{aasjournal}

\end{document}